\documentclass[a4paper]{article}

\usepackage{INTERSPEECH2022}
\usepackage{hyperref}
\usepackage{amsmath,amsfonts}
\usepackage{amssymb}
\usepackage{caption}
\usepackage{subcaption}
\newcommand{\norm}[1]{\left\lVert#1\right\rVert}

\title{Recent improvements of ASR models in the face of adversarial attacks}
\name{Raphael Olivier$^1$, Bhiksha Raj$^1$}
\address{
  $^1$Language Technologies Institute, Carnegie Mellon University, USA}
\email{rolivier@cs.cmu.edu, bhiksha@cmu.edu}

\begin{document}

\maketitle
\begin{abstract}
  Like many other tasks involving neural networks, Speech Recognition models are vulnerable to adversarial attacks. However recent research has pointed out differences between attacks and defenses on ASR models compared to image models. Improving the robustness of ASR models requires a paradigm shift from evaluating attacks on one or a few models to a systemic approach in evaluation. We lay the ground for such research by evaluating on various architectures a representative set of adversarial attacks: targeted and untargeted, optimization and speech processing-based, white-box, black-box and targeted attacks. Our results show that the relative strengths of different attack algorithms vary considerably when changing the model architecture, and that the results of some attacks are not to be blindly trusted. They also indicate that training choices such as self-supervised pretraining may significantly impact robustness by enabling transferable perturbations. We release our source code as a package that should help future research in evaluating their attacks and defenses.
\end{abstract}

\section{Introduction}

In recent years Automatic Speech Recognition (ASR) systems have invested an considerable number of practical applications, and security is a critical component of many of them. Yet there is a dangerous lack of certainty regarding the robustness of ASR models to security threats. For instance neural Speech Recognition systems are vulnerable to adversarial attacks, one of the most emblematic shortcomings of modern Deep Learning architectures \cite{goodfellow13,goodfellow2014}. These attacks lead models to make wrong predictions by crafting small, hardly perceptible perturbations of their inputs. Applied to domains like banking\cite{banking} or home security\cite{security}, where ASR systems are used, such attacks could lead to critical failures and potentially enable fraudulent transactions or theft.

Adversarial attacks against ASR models have been around for several years and their results on baselines like DeepSpeech2 \cite{deepspeech2} or Speech Commands \cite{speechcommand} are well known \cite{houdini,carlini18,alzantot2018did}. Research on Speech Recognition has since then hit multiple milestones by adopting Transformer models, looking into various architectures and even self-supervision. Yet most works on ASR attacks still often evaluate them on the same baseline models or a very limited number of models \cite{taori19,qin19,Xie2021EnablingFA}. This leaves a number of fundamental questions largely unanswered: how do LSTMs and Transformers compare in terms of robustness? What about encoder-decoders and encoder-only models? Does self-supervised pretraining, a staple of recent ASR systems, affect adversarial vulnerabilities? Robustness for Computer Vision has seen tremendous progress in recent years, largely thanks to standardized benchmarks\cite{carlini16,croce2021robustbench} and baselines \cite{madry18,cohen19}. At the same time Randomized Smoothing is the only strong, principled defense that has been adapted to Speech Recognition \cite{olivier-raj-2021-sequential}. Recent works \cite{abdullah2020sok,abdullah2022demystifying} have shown that ASR adversarial examples are different from image ones in key aspects, and must therefore be the object of specific research. In that regard, setting the foundations for a standardized, systematic evaluation of attacks on speech models is an important first step.

Our contribution throughout this paper consists in laying out some of these foundations. We start by reproducing a representative panel of attacks: targeted and untargeted, with and without labels, white-box and black-box, optimization and signal processing-based. We evaluate them on the LibriSpeech test-clean set against multiple standard models trained on the same data but varying in architecture, loss, training paradigm and size. These results both set a standard to beat in future attack and defense works, and let us draw some (cautious) conclusions on the relative robustness of ASR architectures.

We also investigate the conditions under which adversarial examples may transfer between models. We study multiple architectures - notably variations of Wav2Vec2 that were pretrained with self-supervised objectives. Surprisingly, we show that the transferability of adversarial examples is significantly boosted for such models, even for targeted attacks - which so far were systematically non-transferable between ASR models \cite{abdullah2022demystifying}. To an extent, these results indicate that ASR models today are \textit{more} vulnerable than in the past. We draw attention on the dangers of fine-tuning private models from publicly available speech embeddings.

Arguably one of the limitations for a standard evaluation of ASR attacks until this point was the lack of a practical codebase to do so. Our final contribution consists in releasing our source code\footnote{\url{https://github.com/RaphaelOlivier/robust\_speech}} in the form of an simply extensible package built on the Speechbrain toolkit \cite{speechbrain}. We hope this will help performing a thorough evaluation of adversarial attacks in the future.

\section{Adversarial attacks on ASR}
In this section we describe the attacks that we evaluated. We adopt the taxonomy of \cite{abdullah2020sok}.

\subsection{Untargeted attacks}
\subsubsection{Projected Gradient Descent}
The Projected Gradient Descent (PGD) attack is an direct optimization attack. We use the untargeted PGD, which aims at producing denial of service by by generating any wrong output. It optimizes the following objective with projected gradient descent for multiple steps:
\begin{equation}\label{eqn:pgd}
    argmax_{\norm{\delta}<\epsilon}L(f(x+\delta),y)
\end{equation}
where $f$ is the model, $L$ its loss function, $(x,y)$ the original input and label, and $\epsilon$ a bound on the adversarial perturbation $\delta$. PGD is a general machine learning attack and has been applied to ASR models in the past \cite{abdullah2020sok,olivier-raj-2021-sequential}. We consider both the $L_\infty$ and $L_2$ for the choice of norm in Eq. \ref{eqn:pgd}.  We control the perturbation size of the latter by fixing the Signal-Noise Ratio (SNR) of the attack as a hyperparameter. The SNR is defined with $$SNR(\delta,x)=log_{10}(20\frac{\norm{x}_2}{\norm{\delta}_2})$$ and measured in decibels (dB). For each input we set $\epsilon$ accordingly: $\epsilon=\norm{x}_2/10^{\frac{SNR}{20}}$. For $L_\infty$ PGD, we directly set $\epsilon$ as a hyperparameter.

\subsubsection{Genetic ASR attack}
\cite{alzantot2018did} propose an indirect optimization attack for ASR models. It employs a genetic algorithm where a population is maintained and renewed, using the adversarial objective as a fitness score. Such indirect attacks are useful to attack models that are only available as a black-box oracle, without gradient information; but also to fool models whose gradients are not useful due to the \textit{obfuscation} effect of an incomplete defense \cite{athalye18}.

The authors of \cite{alzantot2018did} use this method as a targeted attack. However they only evaluated it on the Speech Commands dataset with 10 possible labels. To run such an attack on a large ASR model with a full english sentence as the target is considerably more challenging. To keep some amount of attack success we use the same algorithm for a simpler, untargeted objective: the loss on the original label is used as a fitness score. Perturbations are $L_\infty$-bounded by a hyperparameter $\epsilon$.

\subsubsection{Kenansville attack}
The DFT Kenansville attack \cite{abdullah2019hear} is a Signal Processing attack: it does not consider model predictions or loss to modify the audio signal. Instead, this attack removes the spectral components of the input that have the lowest power spectral density (PSD). This leaves human perception mostly unchanged, but typically affects the predictions of ASR models, which tend to exploit high-frequency components more than humans. Like $L_2$-PGD the distortion of that attack is controlled with the SNR.
\subsection{Targeted attacks}
\subsubsection{Carlini\&Wagner}
The Carlini\&Wagner (CW) attack \cite{carlini16} is a targeted, direct optimization attack and a standard for evaluating the robustness of machine learning models. The same authors apply it to the DeepSpeech2 ASR model in \cite{carlini18}. At its core is the optimization of $\delta$ to minimize the objective $$c.L(f(x+\delta),y_t) + \norm{\delta}_2^2 $$ such that $\norm{\delta}_\infty<\epsilon$. Here $y_t$ is the target of the attack and $c$ a regularization parameter controlling the relative importance of the attack objective and the size of the perturbation. The objective is optimized with Adam \cite{adam}. The additional $L_\infty$ constraint $\epsilon$ is first set large, then decreased as the attack achieves its objective.

\subsubsection{Imperceptible attack}
The Imperceptible attack \cite{qin19} extends the CW attack by adding a second attack stage where the $\norm{\delta}_2^2$ regularization is replaced with a regularization term based on a psychoacoustic model. This leads to less perceptible adversarial examples. It is also a rare example of attack proposed against a LAS-type model \cite{Chan16LAS} rather than DeepSpeech. We do not report metrics for this attack (they are nearly identical to CW), but we discuss it and provide adversarial examples along with our source code.

\subsection{Self-supervised attack}
Recent years have seen the emergence of self-supervised pretraining for ASR models such as Wav2Vec 2.0 \cite{schneider19_interspeech,Baevski20W2V}. These models are pretrained on unlabeled speech data. During pretraining, given input $x$ the network computes both a quantized latent representation $q(x)$ from low-level features and a contextualized representation $c(x)$ with a Transformer. It is trained to match these two vectors with a contrastive loss.

An adversarial example for such a model would intuitively apply a small perturbation $\delta$ that makes the representation of $c(x+\delta)$ very different from $c(x)$. Even in absence of a training label, attacking self-supervised models is possible \cite{Kim20ASS}. To attack Wav2Vec2 we wrote an algorithm inspired by \cite{sslspeechadv}, in which our loss term is simply the squared distance of the natural and adversarial representations $\norm{c(x+\delta)-c(x)}^2_2$. By plugging this loss term into the $L_\infty$-PGD objective, we obtain the  "untargeted" \textit{SSL attack} on Speech.

\subsection{Transferred attacks}
\cite{papernot16} have shown that adversarial examples computed with optimization attacks against a model tend to be effective at fooling other models as well. This \textit{transferability} effect is a major cause on concern: by attacking a source, proxy model that was trained locally or publicly available, and then transferring the perturbations to the actually targeted model, an adversary can fool even commercial models of unknown architecture.

However, transferability is rarely claimed by attacks on ASR models. In fact \cite{abdullah2020sok} have shown that it is not achievable with current targeted optimization attacks on the DeepSpeech2 model, even when the models differ only by their random training seed. \cite{abdullah2022demystifying} analyse this phenomenon on a Speech Command model and identify factors limiting transferability, such as recurrent networks or vocabulary size. We extend this analysis with a wider set of architectures and by studying untargeted attacks as well. We also study the effect of self-supervised pretaining on transferability.

\section{Experimental setup}
\subsection{Dataset}
We use the LibriSpeech dataset \cite{librispeech}. All our models are trained on the full 960h training set, with the exception of the wav2vec2-100h model which was pretrained on the full set and fine-tuned on a 100h clean subset.

We evaluate our adversarial attacks on the clean test set. For the most computationally expensive attacks, CW, Imperceptible and Genetic, we restrict our evaluation to 100 utterances only.

\subsection{Models}
In order to minimize the time and the computational footprint of our experiments, we evaluate pretrained models directly provided by SpeechBrain \cite{speechbrain}. Some models output characters (31 output neurons), others output subwords with 5000 Byte Pair Encodings (BPE). All subwords (resp. character) models share their vocabulary and tokenizer. We do not use external language models for decoding. The detailed hyperparameters of all models are available in our source code.

We use two subword Seq2Seq encoder-decoder models following the Listen, Attend and Spell (LAS) architecture \cite{Chan16LAS}. One uses Transformers and and the other LSTMs. Each was trained with both the CTC Loss on the encoded states and Attention+Cross-Entropy Loss. We use only the Cross-Entropy loss and Seq2Seq decoding to run and evaluate attacks on LAS models. 

By isolating each encoder and applying CTC decoding, we obtain two encoder-only ASR models that achieve good results. We treat these encoders as separate models, sharing their latent speech space with the Seq2Seq models.

%We train an additional LSTM CTC model, this time with character outputs rather than subwords, in order to better analyse the impact of the network's output size on adversarial vulnerabilities.

Finally, we use multiple variations of Wav2Vec 2.0: the base pretrained model, base models trained on 960h and 100h respectively, and the large model trained on 960h. All models were imported using the Speechbrain interface to Huggingface transformers \cite{wolf-etal-2020-transformers}. %For each of these models, in order to align the output space with our character vocabulary, the final projection was briefly fine-tuned on 100h for one epoch. This was sufficient to achieve the expected performance of these models on LibriSpeech.

\subsection{Evaluation}
\subsubsection{Attack methodology}
We evaluate all attacks against all models whenever applicable. For the PGD and Kenansville attacks we vary the hyperparameters controlling the amount of noise distortion; however for each attack we keep all other hyperparameters constant. This includes the number of iterations in optimization attacks, which is respectively equal to 100 (PGD), 500 (SSL), 2000 (Genetic) and 5000 (CW). These values are greater or equal to those used by their original authors. Our targeted attacks use a set of three possible target sentences. For each input we select the target the closest length to the true label.
\subsubsection{Metrics}
To evaluate our attacks we rely on several metrics:
\begin{itemize}
    \item For untargeted attacks, the \textbf{Word-Error Rate} (WER) of the model on the adversarial example with respect to the \textit{true label}. The higher, the stronger the attack.
    \item For targeted attacks, the WER on the adversarial example with respect to the \textit{target sentence}. The lower, the stronger the attack. We also report the \textbf{accuracy}, or success rate of these attacks, i.e. the fraction of inputs for which a successful adversarial example was found.
    \item The \textbf{Signal-Noise Ratio} (SNR) of the perturbation: the higher the SNR, the less perceptible the noise. $20dB$ can be considered a threshold for "acceptable" noise level that does not disturb human comprehension; $30-40dB$ is an estimate of "imperceptible" noise. Both these threshold are approximate, as human perception does not align well with $L_p$ norms.
\end{itemize}

Since we vary the bounds of PGD and Kenansville we can, for these attacks, plot the WER as a function of the SNR.
\section{Results and discussion}
\begin{table}[h!]
\centering
\begin{tabular}{|c|c|ccc|c|}
 \hline
 Model & Clean & \multicolumn{3}{|c|}{Carlini\&Wagner}& Genetic  \\
 &WER&WER$\downarrow$&SNR&Acc. &WER$\uparrow$ \\
 \hline
 %LSTM LAS 1000  & 4.69 & 9.71 & 18.83 & 17 & 74\% \\
 LSTM LAS  & 5.02 & 16.98 & 14 & 66\%   & 25.49\\
 LSTM CTC  & 6.15 & 4.63 & 23 & 91\% & 24.27  \\
 %LSTM CTC Char  &  &  &  &  &   \\
 Trf LAS & 4.11& 0 & 37 & 100\% & 21.6   \\
 Trf CTC & 5.88 & 0 & 40 & 100\%  & 20.63  \\
 W2V-100h & 6.27 & 0 & 35 & 100\%  & 13.19  \\
 W2V-960h & 3.53 & 0 & 35 & 100\%  & 16.63  \\
 W2V-large & 2.85 & 0 & 32 & 100\%  & 11.08  \\
 \hline
\end{tabular}
\caption{results of all models on clean data and under the (targeted) Carlini\&Wagner attack and the (untargeted) Genetic attack. For CW we report the WER, SNR and attack accuracy (i.e. success rate). The bound for the Genetic attack corresponds to an average SNR of $20dB$. The up and down arrows indicate whether large or small WER values mean successful attacks.}
\label{table:attack}
\end{table}
\begin{figure*}[h]
     \centering
     \begin{subfigure}[b]{0.29\textwidth}
         \centering
         \includegraphics[width=\textwidth]{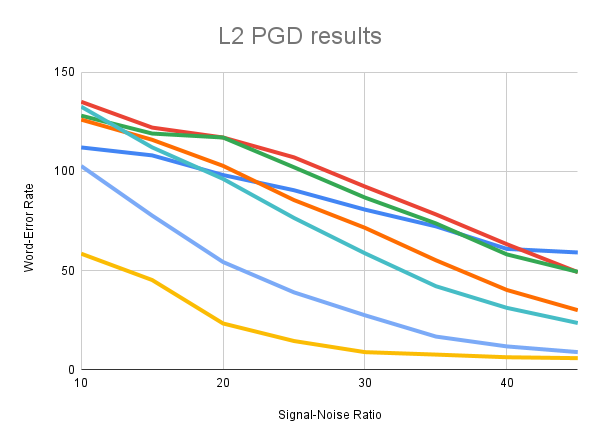}
         \caption{WER as a function of SNR for the $L_2$-PGD attack}
         \label{fig:l2pgd}
     \end{subfigure}
     \hfill
     \begin{subfigure}[b]{0.29\textwidth}
         \centering
         \includegraphics[width=\textwidth]{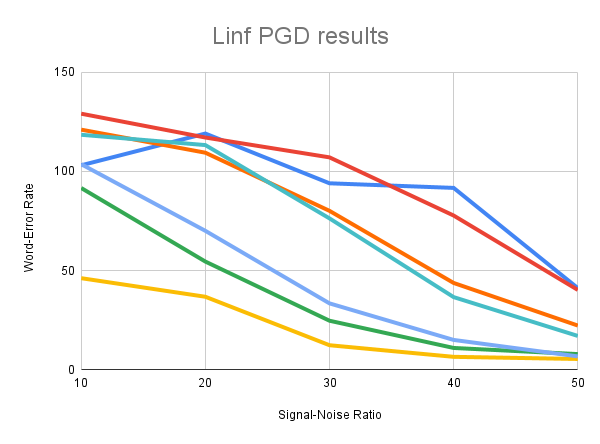}
         \caption{WER as a function of SNR for the $L_\infty$-PGD attack}
         \label{fig:linfpgd}
     \end{subfigure}
     \hfill
     \begin{subfigure}[b]{0.36\textwidth}
         \centering
         \includegraphics[width=\textwidth]{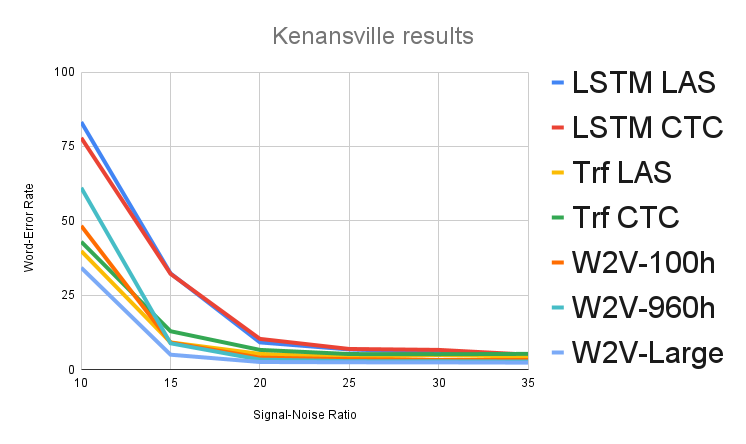}
         \caption{WER as a function of SNR for the DFT Kenansville attack}
         \label{fig:kenansville}
     \end{subfigure}
     \hspace*{\fill}
        \caption{Attacks results for the untargeted (a) $L_2$-PGD attack, (b) $L_\infty$-PGD attack and (c) DFT Kenansville attack, for different SNR values. The legend on the right is shared by all plots.}
        \label{fig:three graphs}
\end{figure*}
We plot the results of the PGD and Kenansville attacks in Figure \ref{fig:three graphs}, while the CW and Genetic attack results are displayed in Table \ref{table:attack}. 

\subsection{Direct optimization attacks}

We start by analysing direct optimization attacks: PGD (Fig \ref{fig:l2pgd} , \ref{fig:linfpgd}), and  CW (Table \ref{table:attack} columns 2-5). 

The analysis of the PGD plots show large variations in the WER of different models under attack. The relative ranks of all models are consistent between $L_\infty$ and $L_2$ attacks. The only exception is the Transformer CTC encoder, which compared to other models is more robust to $L_\infty$ attacks than $L_2$ attacks. CW results on the other hand show a clear gap between the Transformer models and the LSTM models: the former always predict the attack target on adversarial examples with low noise ($SNR\geq30dB$), while the latter require more noise for imperfect target transcriptions. The Imperceptible attack achieves almost identical results (not reported), with SNRs varying by up to 3dB.

Overall two trends emerge from these results: 

\begin{itemize}
    \item Encoders-decoder are \textit{harder} to attack than encoder-only models. LAS models get lower WER than CTC models on PGD and lower SNR on CW attacks. The Transformer LAS model has a lower PGD WER than the Transformer encoder.
    \item Transformers are \textit{harder} to attack than LSTMs with PGD, but \textit{easier} with CW.
\end{itemize}

The higher robustness of encoder-decoders may simply be a result of their higher depth, which is a known cause of gradient obfuscation \cite{athalye18}. The fact that CW is effective against Transformer Seq2Seq models points towards that explanation: this attack's objective is especially powerful at circumventing optimization issues. The inconsistent results of PGD and CW against LSTMs and Transformers are harder to explain, and certainly deserve more attention in future works. At the very least it shows that the performance boost of transformers does not clearly carry over in terms of robustness. It also emphasizes the need to evaluate multiple models when crafting new attacks, and multiple attacks when building new models.

\subsection{Other attacks}

The Kenansville DFT attack (Figure \ref{fig:kenansville}) has little effect in our evaluation setting: SNRs greater than 20dB fail to increase the WER significantly, and adversarial examples with lower SNRs are easily detected by the human ear. Comparing these results to \cite{abdullah2019hear} is not easy for two reasons: the authors report different metrics to measure distortion, and evaluate on a different dataset. The adversarial examples they publicly released show that the original speech inputs are of lower quality than the LibriSpeech clean test set. It is therefore possible that Kenansville is most effective on inputs that are already challenging to the ASR models.

The Genetic attack \cite{alzantot2018did} (Table \ref{table:attack} column 6) also only achieves to degrade the WER of each model by 8 to 20 points, even with an SNR of 20dB. This is not particularly surprising: this attack was originally evaluated on the much less challenging Speech Commands dataset.

\subsection{Transferred attacks}

\begin{table}[h!]
\centering
\begin{tabular}{|c|c|c|c|} 
 \hline
 Source Model & Target model & PGD & CW \\ [0.5ex] 
  &  & WER$\uparrow$ & WER$\downarrow$ \\ [0.5ex] 
 \hline
 LSTM CTC & LSTM LAS & \textit{121} & 85.8  \\
 Trf CTC & LSTM LAS & 14.13 & 95.83 \\
 LSTM CTC & Trf LAS  & 34.28 & 108 \\
 \hline
 W2V-960h & W2V-100h  & \textit{80.01} &  \textit{48.15}  \\
 W2V-100h & W2V-960h & \textit{84.38} & \textit{61.11} \\
 W2V-960h & W2V-large  & 45.89 & 91.4  \\
 W2V-large & W2V-960h & 37.46 & \textit{62.65}  \\
 \hline
\end{tabular}
\caption{Transferred attack results on selected source/target model pairs. The $L_2$-PGD SNR bound is 25dB. Results in italic indicate full or partial attack transferability.}
\label{table:transferred}
\end{table}
\begin{table}[h!]
\centering
\begin{tabular}{|c|c|c|c|} 
 \hline
 Target model & WER$\uparrow$ ($30dB$) & WER$\uparrow$ ($20dB$) \\ [0.5ex] 
 \hline
  LSTM LAS  & 6.89 & 23.48  \\
  LSTM CTC  & 7.91 & 23.23   \\
  %LSTM CTC Char  &  &   \\
 Trf LAS & 4.13 & 8.43  \\
  Trf CTC & 6.31 & 12.85  \\
W2V-100h & 28.64 & \textit{82.28}  \\
  W2V-960h & 48.79 & \textit{96.36} \\
  W2V-large  & 3.4 & 9.95 \\
 \hline
\end{tabular}
\caption{SSL attack results on all models, with SNR bounds of $30dB$ and $20dB$. Attacks are transferred from the Wav2Vec2-base pretrained model. Results in italic indicate full or partial attack transferability.}
\label{table:ssl}
\end{table}
In Table \ref{table:transferred} we report the results of transferred PGD and CW attacks between selected source-target model pairs. 

The first three rows report our results on supervised LAS and CTC models. They confirm the results that \cite{abdullah2020sok} achieved on DeepSpeech2: the CW attack displays almost no transferability between ASR models ($\text{WER}\geq90\%$), for example between Transformer encoder and LSTM Seq2Seq. Untargeted PGD attacks achieve only a partial degradation of target model's performance, even at 25dB SNR. The only exception is the PGD transfer between LSTM CTC and LAS models, i.e. models that share the same encoder. And even such parameter sharing is not sufficient to transfer CW perturbations ($\text{WER}\geq85\%$).

However, results are different between Wav2Vec2 models. When transferring perturbations between Wav2Vec2 base models fine-tuned from the same pretrained encoder, we achieve a PGD degradation to 80\% WER, and even partial transferability for CW (48\% and 61.11\%). These results are confirmed by the SSL attack (Table \ref{table:ssl}): perturbations crafted against the Wav2Vec2 base pretrained encoder substantially degrade the performance of both Wav2Vec2 fine-tuned models, but have very limited effect on any other model. With CW we even observe partial transferability from Wav2Vec2-Large to Wav2Vec2-base ($\text{WER}=62.65$), while these models have different architectures and were not using the same initial weights. This shows that part of this transferability is indeed due to the Wav2Vec2 architecture and pretraining, and not just to weight initialization.

To our knowledge, this is the first case of targeted \textit{and} transferable adversarial attack on ASR models. These findings are particularly worrisome given the recent trend in Natural Language Processing, consisting in training a few large self-supervised models that are applied to many contexts with small amounts of fine-tuning or even zero-shot prompting. Should Speech Recognition take a similar path, ASR adversarial attacks may become a much greater concern.

\section{Conclusion}
In this work we have shown the relative strength of Transformers vs LSTMs and LAS vs CTC models, with respect to adversarial attacks. We have updated the state of multiple standard attacks against recent ASR models. Finally, we have illustrated the surprising vulnerability of Wav2Vec2 to transferred targeted attacks compared to all other ASR architectures studied in the past. Our results hint towards self-supervised pretraining as a condition for the transferability of ASR adversarial examples.

Direct extensions of this work involve evaluating more attacks, like stronger Black box attacks \cite{taori19}; and applying attacks to Transducer ASR models, which are popular in streaming Speech Recognition applications \cite{rnnt}.\footnote{Since no pretrained Transducer is currently provided with Speechbrain, these extensions require important computational efforts.}. More generally, we hope that our methodology, findings and code will help accelerate research efforts on ASR robustness.
\bibliographystyle{IEEEtran}

\bibliography{template}

% \begin{thebibliography}{9}
% \bibitem[1]{Davis80-COP}
%   S.\ B.\ Davis and P.\ Mermelstein,
%   ``Comparison of parametric representation for monosyllabic word recognition in continuously spoken sentences,''
%   \textit{IEEE Transactions on Acoustics, Speech and Signal Processing}, vol.~28, no.~4, pp.~357--366, 1980.
% \bibitem[2]{Rabiner89-ATO}
%   L.\ R.\ Rabiner,
%   ``A tutorial on hidden Markov models and selected applications in speech recognition,''
%   \textit{Proceedings of the IEEE}, vol.~77, no.~2, pp.~257-286, 1989.
% \bibitem[3]{Hastie09-TEO}
%   T.\ Hastie, R.\ Tibshirani, and J.\ Friedman,
%   \textit{The Elements of Statistical Learning -- Data Mining, Inference, and Prediction}.
%   New York: Springer, 2009.
% \bibitem[4]{YourName17-XXX}
%   F.\ Lastname1, F.\ Lastname2, and F.\ Lastname3,
%   ``Title of your INTERSPEECH 2022 publication,''
%   in \textit{Interspeech 2022 -- 23\textsuperscript{rd} Annual Conference of the International Speech Communication Association, September 18-22, Incheon, Korea, Proceedings, Proceedings}, 2022, pp.~100--104.
% \end{thebibliography}

\end{document}